\documentclass[10pt,superscriptaddress,twocolumn,amsmath,amssymb,aps,pra]{revtex4}
\usepackage{epstopdf}
\usepackage{mathrsfs}
\usepackage{graphicx}% Include figure files
\usepackage{dcolumn}% Align table columns on decimal point
\usepackage{bm}% bold math
\usepackage[utf8]{inputenc}
\newcommand{\be}{\begin{equation}}
\newcommand{\ee}{\end{equation}}
\newcommand{\bea}{\begin{eqnarray}}
\newcommand{\eea}{\end{eqnarray}}

\begin{document}
\title{The Atomic Superfluid Quantum Interference Device with tunable Josephson Junctions}
\author{Jiatao Tan}
\affiliation{Institute of Theoretical Physics, School of Physics and Optoelectronic Engineering, Beijing University of Technology, Beijing, 100124, China}
\author{Boyang Liu}\email{boyangleo@gmail.com}
\affiliation{Institute of Theoretical Physics, School of Physics and Optoelectronic Engineering, Beijing University of Technology, Beijing, 100124, China}

\date{\today}
\begin{abstract}
The atomic superfluid quantum interference device (ASQUID) with tunable Josephson junctions is theoretically investigated. ASQUID is a device that can be used for the detection of rotation. In this work we establish an analytical theory for the ASQUID using the tunneling Hamiltonian method and find two physical quantities that can be used for the rotation sensing. The first one is the critical population bias, which characterizes the transition between the self-trapping and the Josephson oscillation regimes and demonstrates a periodic modulation behavior due to the rotation of the system. We discuss the variation of critical population bias when the tunneling strengths of the junctions are tuned in different values, and find that the symmetric junctions are better choice than the asymmetric ones in terms of rotation sensing. Furthermore, how the initial phase difference between the two condensates affects the measurement of rotation is also discussed. Finally, we investigate the case of time-dependent junctions and find there is another physical quantity, named critical time, that can be used to detect the rotation.
\end{abstract}
 \maketitle
\section{Introduction}
Atomtronics\cite{Amico2021,Amico2022} is an emerging field in quantum technology, focusing on the manipulation and control of ultra-cold atoms to create devices analogous to conventional electronics. Atomtronics utilizes neutral atoms in optical lattices or magnetic traps to form circuits, devices, and systems with unique quantum properties. In the field of quantum sensing
%Macroscopic quantum systems such as superconductors, superfluids exhibit fascinating transport properties. One of the most interesting phenomena is the Josephson effect between two weakly coupled superconductors. This phenomenon can be employed in interferometer devices for quantum sensing,
the SQUID is widely used for measuring extremely weak magnetic fields \cite{Clarke2004}. Inspired by the  applications of SQUIDs, the development of their atomic counterparts, ASQUIDs, attracts a lot of attention and opens up new possibilities for precision measurement \cite{Wright2013,Ryu2013,Jendrzejewski2014,Eckel2014,Ryu2020}. The major difference between SQUID and ASQUID is that the superfluids in ASQUIDs are made of neutral atoms instead of Cooper pairs. Therefore, the ASQUID doesn't couple to the electro-magnetic fields. The phase twist of the condensates in ASQUID can be created by the physical rotation of the device, so ASQUID can be used for rotation sensing.
With the advantages of high controllability and detection of atom number and phase of condensates, ASQUID can be a good platform for studying basic properties of quantum systems and motion sensing.

\begin{figure}[h]
\includegraphics[width=0.35\textwidth]{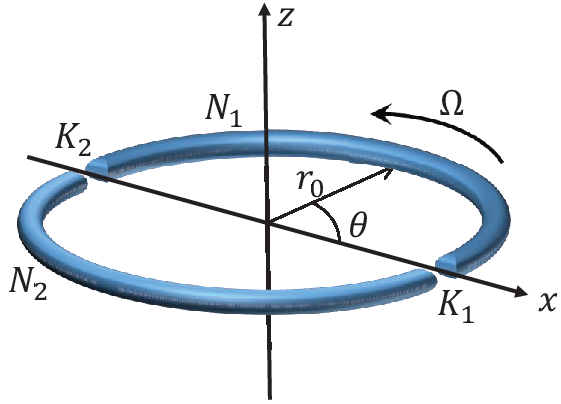}
\caption{(Color online) The sketch of the ASQUID with double junctions. The tunneling strength of the junctions are denoted by $K_1$ and $K_2$. The particle numbers in the two regions are $N_1$ and $N_2$. The ASQUID is rotating with an angular velocity $\Omega$. }
\label{fig:ASQUID}
\end{figure}

In the past decades, a number of ring-shaped traps have been realized based on magnetic and optical dipole trapping \cite{Gupta2005,Arnold2006,Ryu2007,Ramanathan2011,Cai2022,Pace2022}, which paves the way for the creation of ASQUIDs. As sketched in Fig.\ref{fig:ASQUID} a typical ASQUID is formed by confining Bose-Einstein condensate (BEC) in a ring-shaped trap. Potential barriers acting as Josephson junctions are created by blue-detuned laser beams. In Fig. \ref{fig:ASQUID} the two junctions are located at $\theta=0$ and $\theta=\pi$, and they divide the ring-shaped trap into two regions. The potential barriers can be set into motion to create particle number density bias or simulate the rotation of the system\cite{Jendrzejewski2014,Ryu2020}. ASQUIDs with single junction and double junctions have been constructed and studied in the cold-atom experiments, simulating the radio frequency (RF) SQUIDs and the direct current (DC) SQUIDs in the traditional electronics, respectively. In the setup with single junction the phase slips \cite{Wright2013} and hysteresis \cite{Eckel2014} have been observed. In the one with double junctions the Josephson effect \cite{Ryu2013} and resistive flow \cite{Jendrzejewski2014} have been investigated, and interestingly the response of such a system subject to rotation was observed recently \cite{Ryu2020}.

The usual theoretical analysis on the ASQUID is to directly implement the DC SQUID theory that was employed in the conventional SQUID, however, which is not a complete theory for the ASQUID.  One of the major differences between the conventional SQUID and the ASQUID is that the conventional SQUID is connected to a battery, and hence a constant voltage drop can be maintained across the SQUID. The particle number bias in the SQUID is then a constant and a steady current through the SQUID can be derived. The ASQUID is an isolated system. In this case the particle numbers in each region vary with time. Hence, a theory suitable for the ASQUID has to be established. In this work we systematically construct a theory that describes the setup of ASQUID depicted in Fig. 1. Based on this theory we found two physical quantities that can be used for sensing the rotation, and we discussed how the variation of the Josephson junctions affects the motion sensing.

Our work is organized as the following. In Sec. \ref{sec:model} we construct a model for the setup in Fig. \ref{fig:ASQUID} using the tunneling Hamiltonian method, and based on which we obtain the differential equations that describe the dynamics of the system. In Sec. \ref{sec:ZcS} we introduce a physical quantity that can be used for the rotation sensing, the critical population bias. We study the response of it to the rotation in case of symmetric junctions.  In Sec. \ref{sec:ZcA} we study the behavior of the critical population bias in case of asymmetric junction. In Sec. \ref{sec:tc} the case of dynamical Josephson junctions is investigated. Here, we introduce another physical quantity that can be used for the rotation sensing, namely, the critical time. Finally, Sec. \ref{sec:con} provides our conclusions.
\section{Model\label{sec:model}}
We consider a model with BEC confined in a quasi-1D ring-shaped trap as sketched in Fig. \ref{fig:ASQUID}. Using the tunneling Hamiltonian method the Hamiltonian of the system can be cast into three parts as the following
\bea
\hat H=\hat H_1+\hat H_2+H_t,
\eea
where
\bea
&&\hat H_{1}=\int_{0}^{\pi}d\theta r_0\Big(-\cfrac{\hbar^2}{2mr_0^{2}}\hat\psi_{1}^\dagger(\theta)\cfrac{\partial^{2}}{\partial\theta^{2}}\hat \psi_{1}(\theta)+U|\hat\psi_{1}(\theta)|^{4}\Big),\cr&&
\hat H_{2}=\int_{-\pi}^{0}d\theta r_0\Big(-\cfrac{\hbar^2}{2mr_0^{2}}\hat\psi_{2}^\dagger(\theta)\cfrac{\partial^{2}}{\partial\theta^{2}}\hat \psi_{2}(\theta)+U|\hat\psi_{2}(\theta)|^{4}\Big).\cr&&
\eea
Hamiltonian $\hat H_i (i=1,2)$  describes the BECs in the $i$-th region, and $\hat \psi^\dagger_i$ and $\hat \psi_i$ are the creation and annihilation operators of the bosons in these regions. $U$ is the interaction strength of the bosons. The particle numbers for region 1 and 2 are $N_1$ and $N_2$, respectively, and the total particle number $N=N_1+N_2$ is fixed. The radius of the quasi-1D ring-shaped trap is $r_0$. The part $\hat H_t$ describes the particle tunneling between the two regions as the following
\bea &&\hat H_{t}=\int_{-\pi}^{\pi}d\theta r_0\Big\{K_{1}\delta(\theta)[\hat\psi_{1}^\dagger(\theta)\hat\psi_{2}(\theta)+\hat\psi_{2}^\dagger(\theta)\hat\psi_{1}(\theta)]\cr&&
~~~~~~+K_{2}\delta(\theta-\pi)[\hat\psi_{1}^\dagger(\theta)\hat\psi_{2}(\theta)+\hat\psi_{2}^\dagger(\theta)\hat\psi_{1}(\theta)]\Big\}.
\eea
$H_t$ describes the tunneling of the bosons through the two Josephson junctions. The locations of the junctions are at $\theta=0$  and $\pi$, and the tunneling strength are described by the parameters $K_1$ and $K_2$, respectively.

Here we consider a rotating system, which can be simulated by the movement of the Josephson junctions in the realistic experiments. In the rotating frame the Hamiltonian is given by
\bea
H^\prime=H-\Omega J_z,
\eea where $\Omega$ is the angular velocity. We are dealing with a uniform rotating system in this work, hence, $\Omega$ is a constant. $J_z=-i\hbar\frac{\partial}{\partial\theta}$ is the angular momentum operator in $\hat z$ direction. When the tunneling between the BECs in the two regions is turned off, the ground states are supercurrent states, which carry angular momenta. The macroscopic wave functions of these states can be written as $\psi_i=\sqrt{n_i}e^{i (m_i\theta+\alpha_i)},(i=1,2)$. $n_i$ is the number densities, and the phases of condensates are linearly dependent on the angle $\theta$. When there is no tunneling, the parameters $n_i$, $m_i$ and $\alpha_i$ are all constants. When the tunneling is turned on, the system becomes dynamical. Since we only consider the case of weak link between the two condensates, we assume the form of the ground state wave function doesn't change except the parameters $n_i$, $m_i$ and $\alpha_i$ become time dependent, which can be expressed as the following
\bea
\psi_i(\theta,t)=\sqrt{n_i(t)}e^{i(m_i(t)\theta+\alpha_i(t))}.\label{eq:WF}
\eea

To study the dynamical properties we employ the action principle $\delta\int^{t_2}_{t_1} L dt=0$, where the Lagrangian is cast as
\bea
L=\int_{0}^{\pi} d\theta r_{0}  \hat\psi_{1}^\dagger i\hbar \partial_{t}\hat\psi_{1}+\int_{-\pi}^{0}d\theta r_{0} \hat\psi_{2}^\dagger i\hbar\partial_{t}\hat\psi_{2}-H^\prime. \eea
In the mean-field level the Lagrangian can be calculated by replacing the field operators $\hat\psi_i$ with the ground state wave function in Eq. (\ref{eq:WF}). Then the dynamical behaviors of system can be derived from the Euler-Lagrange equations
$
\frac{\partial L}{\partial \lambda}=\frac{\partial}{\partial t}\Big(\frac{\partial L}{\partial \dot{\lambda}}\Big)
$, where $\dot{\lambda}=\partial_t\lambda$ and $\lambda$ represents all the six parameters $n_i(t)$, $m_i(t)$ and $\alpha_i(t)$. Straight forward calculation yields
\begin{align}
& m_{+}=\cfrac{4\Omega}{\Omega_0}+\cfrac{8K_{2}}{\hbar\Omega_0\sqrt{1-Z^{2}}}\sin(\Phi+\cfrac{m_{+}}{2}\pi)+\cfrac{2\pi \partial_{t}Z}{\Omega_0(1-Z^{2})},  \nonumber\\
& m_{-}=-\cfrac{8K_{2}Z}{\hbar\Omega_0\sqrt{1-Z^{2}}}\sin(\Phi+\cfrac{m_{+}}{2}\pi)-\cfrac{2\pi Z\partial_{t}Z}{\Omega_0(1-Z^{2})},  \nonumber\\
& \partial_{t}\Phi=-\cfrac{\Omega_0
 m_{-}}{4}(m_{+}-\cfrac{4\Omega}{\Omega_0})-\cfrac{2\tilde{U}Z}{\hbar}\nonumber \\&+\cfrac{2Z}{\pi\hbar\sqrt{1-Z^{2}}}[K_{+}\cos\Phi\cos(\cfrac{m_{+}}{2}\pi)+K_{-}\sin\Phi\sin(\cfrac{m_{+}}{2}\pi)],  \nonumber\\
 &\partial_{t}Z=\nonumber\\&-\cfrac{2}{\pi\hbar}\sqrt{1-Z^{2}}[K_{+}\sin\Phi\cos(\cfrac{m_{+}}{2}\pi)-K_{-}\cos\Phi\sin(\cfrac{m_{+}}{2}\pi)].
\label{eq:EL}\end{align}
In above equations we redefine some of the parameters as $m_\pm(t)=m_1(t)\pm m_2(t)$, $K_\pm=K_1\pm K_2$, and $\Phi=\alpha_1(t)-\alpha_2(t)+\pi m_+/2$. The population bias between the two region is defined as $Z(t)=(N_1(t)-N_2(t))/N$. The interaction strength is redefined as $\tilde{U}=UN/\pi r_0$. In the ring trap the flow of the superfluid is quantized, that is, the phase of the wave function winds up by $2\pi n$ along the ring trap, where $n$ is an integer. Then, the flow circulation can be calculated as $\kappa=nh/m$ \cite{Annett2004}. The circulation with a unit winding number is $h/m$. Correspondingly, we can define a fundamental rotation rate
\bea\Omega_0=h/mA,\label{eq:omega0}\eea where $A=\pi r_0^2$ is the area of the ring trap. Basically, $\Omega_0$ can be taken as the angular velocity of the superfluid. Furthermore, we can also set a fundamental energy scale for the system, that is, the kinetic energy per atom of the superfluid with circulation $h/m$. It can be calculated as $E_0=\frac{1}{2}mr^2_0\Omega^2_0=\hbar\Omega_0$. In this work $E_0$ will be used as the unit of energy.

As we have discussed previously, one of the differences between the conventional SQUID and the ASQUID is that ASQUID is an isolated system and the particle number bias $Z(t)$ varies with time. However, the result of the conventional SQUID can be treated as a special case of above model, that is, it can be reduced from Eqs. (\ref{eq:EL}) with some assumptions. The current through the ASQUID can be calculated as $I\equiv\frac{1}{2}\frac{\partial}{\partial t}(N_1-N_2)=\frac{N}{2}\frac{\partial Z(t)}{\partial t}$. In order to derive the result of the conventional SQUID, we assume very weak links, $K_1=K_2\ll \hbar\Omega$, and the two regions will not be charged up, that is, we set the terms of $\partial Z/\partial t$ in equations of $m_\pm$ to zeros. Then, it's straightforward to derive the current from the forth equation in Eqs. (\ref{eq:EL}) as the following
\bea
I=I_c\sin\Phi,
\eea where $I_c\simeq\frac{NK_+}{\pi\hbar}\cos(2\pi\Omega/\Omega_0)$. This is exactly the result of a superfluid quantum interference device \cite{Simmonds2001}. It's also in the same form as the current of a conventional SQUID, except that the $\cos(2\pi\Omega/\Omega_0)$ part should be replaced by $\cos(e\mathbf{\Phi}/\hbar)$ in the case of SQUID since the SQUID is coupled to the magnetic field\cite{Feynman1965}, where $\mathbf{\Phi}$ is the magnetic flux through the SQUID.

\section{The Critical Population Bias for Symmetric Josephson Junctions\label{sec:ZcS}}
\begin{figure}[t]
\includegraphics[width=0.45\textwidth]{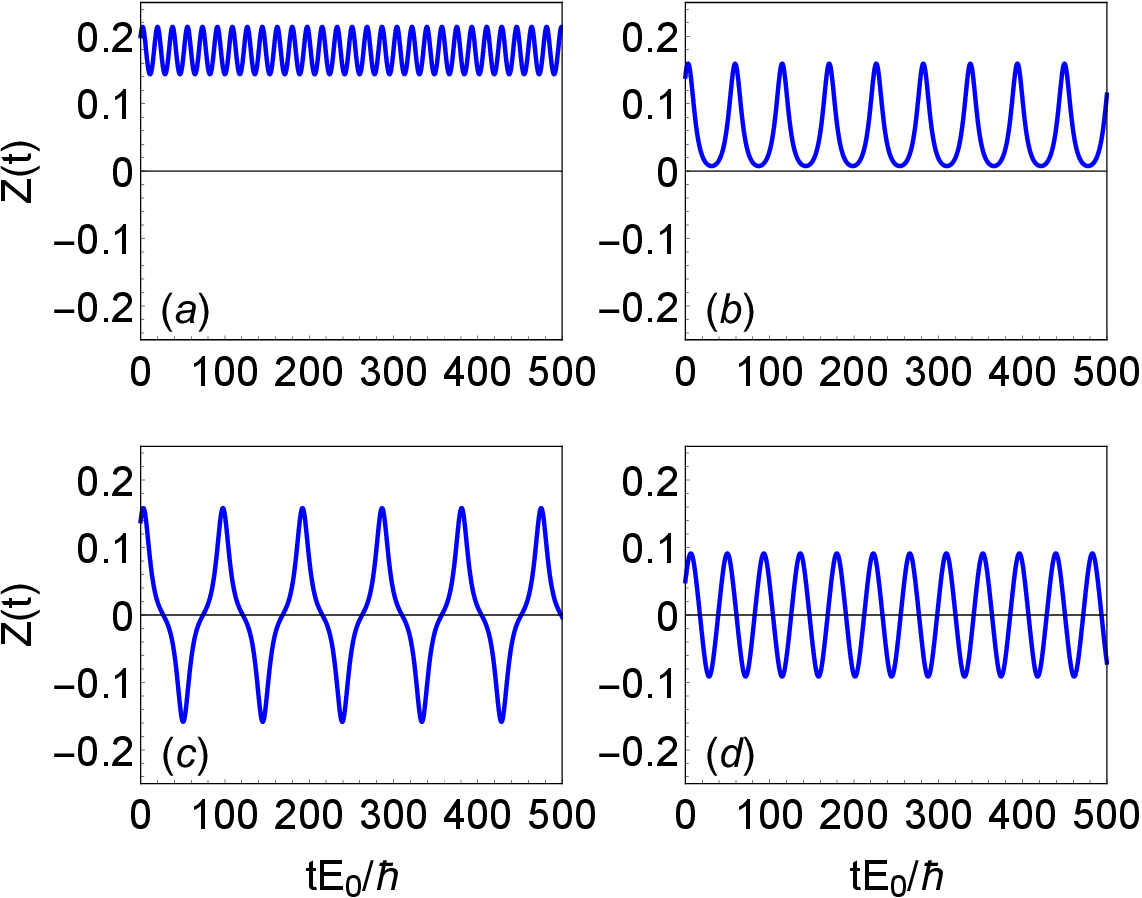}
\caption{(Color online) The temporal variation of the population bias $Z(t)$ for different initial values. (a), $Z(0)=0.2$; (b), $Z(0)=0.14$; (c), $Z(0)=0.139$; (d), $Z(0)=0.05$. The initial phase $\Phi(0)=1$, $\Omega/\Omega_0=0.5$, $K_1/E_0=K_2/E_0=0.01$, $\tilde U/E_0=1$ for all the graphs.}
\label{fig:Zt}
\end{figure}
The conventional SQUID measures the magnetic field by reading the rapid oscillating Josephson current. However, it's difficult to directly observe the Josephson current in ASQUIDs since the ASQUID is an isolated system. Hence, new physical quantities that respond to the rotation has to be found. In this work we find that the critical population bias $Z_c$ demonstrates a periodic modulation behavior due to the rotation. The critical population bias $Z_c$ is the initial population bias where the system switches between the self-trapping regime and the Josephson oscillation regime \cite{Smerzi1997} . In Fig. \ref{fig:Zt} we demonstrate the variation of the population bias $Z(t)$ for different initial values $Z(0)$. For large initial values $Z(0)$ the system shows a behavior of self-trapping as illustrated in Fig. \ref{fig:Zt} (a), while for small $Z(0)$ system shows a behavior of Josephson oscillation as shown in Fig. \ref{fig:Zt} (d). The critical value $Z_c$ for the transition between the self-trapping and the Josephson oscillation is found at the turning point with certain precision. For instance,  when the initial population bias is set to $Z(0)=0.14$  the system is in self-trapping regime as shown in Fig. \ref{fig:Zt} (b). When the initial population bias slightly decreases to $Z(0)=0.139$ the system enters the Josephson oscillation regime as shown in Fig. \ref{fig:Zt} (c). Then the critical population bias is taken as $Zc=0.139$.

\begin{figure}[t]
\includegraphics[width=0.45\textwidth]{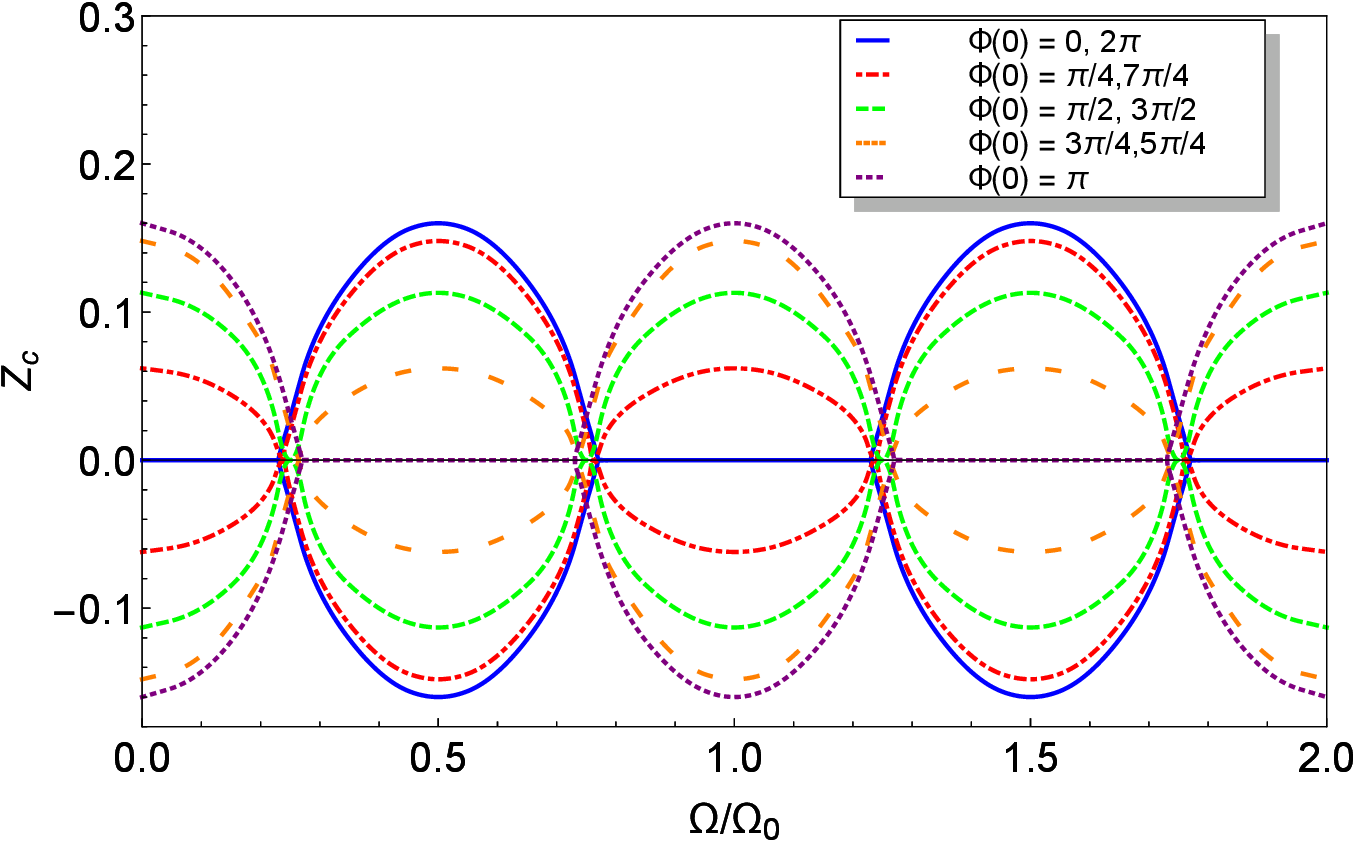}
\caption{(Color online) The critical population bias $Z_c$ as a function of $\Omega/\Omega_0$ for different initial phases $\Phi(0)$. $K_1/E_0=K_2/E_0=0.01$, $\tilde U/E_0=1$ for all the curves.}
\label{fig:ZcS}
\end{figure}

In this section we will focus on the case of symmetric junctions, that is, $K_1=K_2$. We find that the critical population bias $Z_c$ demonstrates a periodic modulation due to the variation of the angular velocity $\Omega$ of the rotation as shown in Fig. \ref{fig:ZcS}. Different curves are corresponding to different initial phase difference $\Phi(0)$. For a fixed phase difference $\Phi(0)$ and $\Omega$ there are two critical population bias with opposite signs since the two reservoirs are symmetric. The periodic modulation of $Z_c$ clearly demonstrate that it can be used to measure the angular velocity of the rotation. The distance between the two adjacent peaks is $\Omega_0/2$, so the sensitivity of the measurement is determined by the quantity of $\Omega_0$. From Eq. (\ref{eq:omega0}) one observes that the sensitivity can be improved by increasing the ring trap radius $r_0$. In the experiment of Ref. \cite{Ryu2020} the fundamental rotation rate $\Omega_0$ is measured. Their definition of $\Omega_0$ is different from ours by a factor of 2. To compare their measurement with our calculations the measured value must be multiplied by 2. After doing that the measured value is roughly $\Omega_0\approx72 $Hz for the case of $r_0=4.82 \mu m$\cite{Ryu2020}. Using Eq.(\ref{eq:omega0}) the $\Omega_0$ is calculated as $\Omega_0=62.6 $Hz, which is close to the measured value.

In cold atom physics the initial phases of the condensates are controllable and can be easily determined by phase imprinting techniques \cite{Zheng2003,Yefsah2013,Kumar2018,Luick2020,Pace2022}. The choice of the initial phase difference $\Phi(0)$ is also important for the measurement of rotation. In Fig. \ref{fig:ZcS} one observes that the curves of $\Phi(0)=0,{\rm or}~\pi$ are totally flat in some regions. For these curves the sensitivity of the measurement of rotation is low since there are no peak structures in some regions. On the other hand, the initial phase close to $\pi/2$ or $3\pi/2$ is better choice for the measurement since they show clear peak structures in all the region.

\section{The Critical Population Bias for Asymmetric Josephson Junctions\label{sec:ZcA}}
\begin{figure}[t]
\includegraphics[width=0.4\textwidth]{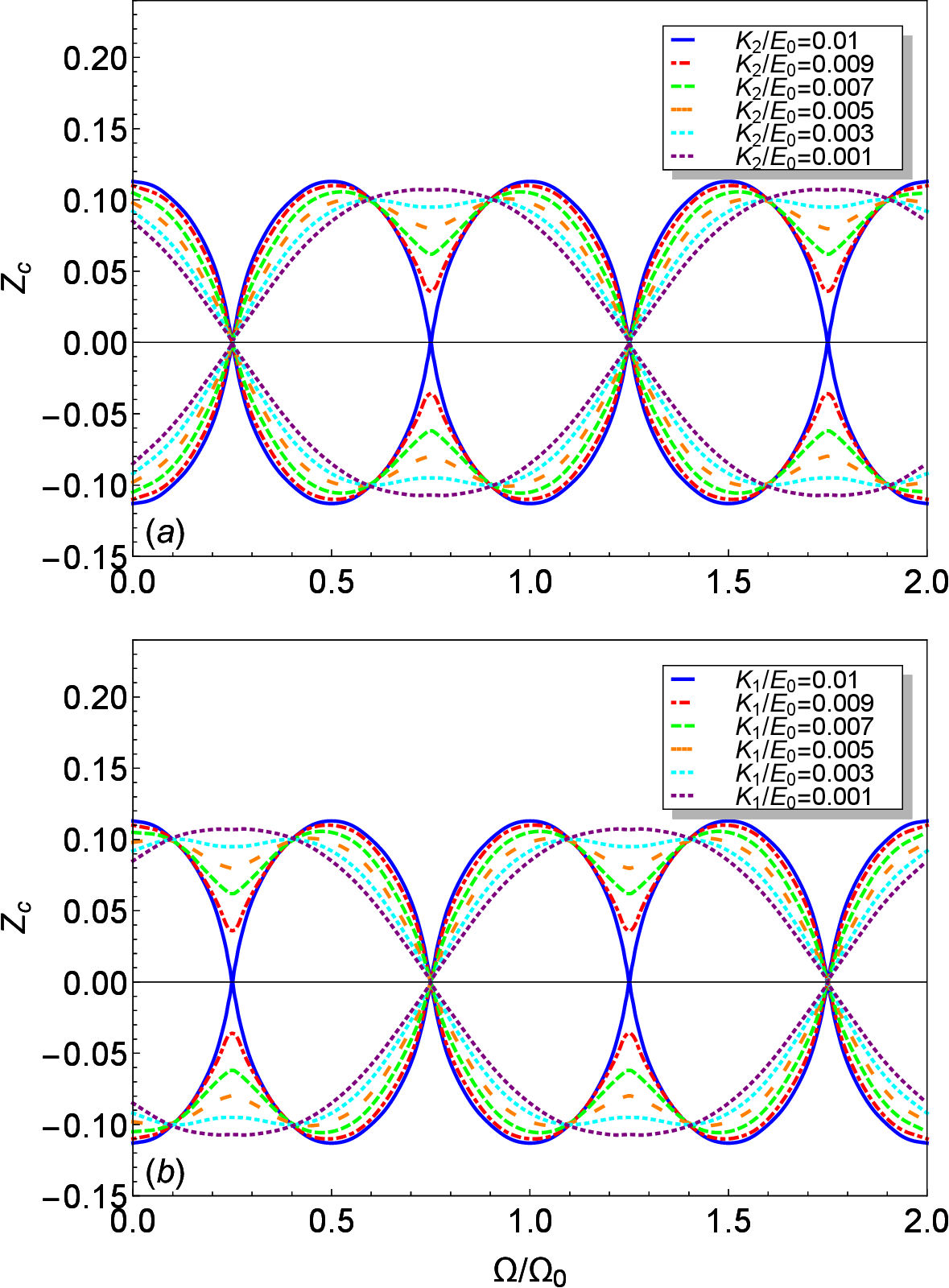}
\caption{(Color online) The critical population bias $Z_c$ as a function of $\Omega/\Omega_0$. In (a) $K_1/E_0$ is fixed at $0.01$, while $K_2/E_0$ descreases from $0.01$ to $0.001$ for different curves.  In (b) $K_2/E_0$ is fixed at $0.01$, while $K_1/E_0$ descreases from $0.01$ to $0.001$ for different curves. The initial phase $\Phi(0)=\pi/2$ and  $\tilde U/E_0=1$ for all the graphs.}
\label{fig:ZcA}
\end{figure}
In cold atom experiments, the Josephson Junctions are created by blue-detuned laser beams, and hence the tunneling strength $K_1$ and $K_2$ can be easily tuned by changing the intensity of the laser beams\cite{Ramanathan2011,Wright2013}. In this section we investigate the effects of the asymmetry of the two junctions, that is, the case of $K_1\neq K_2$. In Fig. \ref{fig:ZcA} (a) we set $K_1=0.01E_0$ and $\Phi(0)=\pi/2$, and plot the critical population bias $Z_c$ for different values of $K_2$, which varies from $0.01E_0$ to $0.001E_0$. When $K_2=K_1$ the curves show an oscillating behavior with period of $\Omega=0.5\Omega_0$. When $K_2$ decreases, the positive and negative branches of the curves start separating from each other and a gap is opened at the points of $\Omega=(n+0.75)\Omega_0,~(n=0,\pm1,\pm2...)$. When $K_2$ is smaller than $K_1$ by one order of magnitude, the period of the curves becomes $\Omega=\Omega_0$. In Fig.\ref{fig:ZcA} (a) we set $K_2=0.01E_0$ and $\Phi(0)=\pi/2$, and plot $Z_c$ for different values of $K_1$. The curves shows similar behaviors except that the positive and negative branches separated from each other at points of $\Omega=(n+0.25)\Omega_0,~(n=0,\pm1,\pm2...)$. From above observations we can conclude that the symmetric junctions are better choice for the rotation sensing since the asymmetry of the junctions can blur two adjacent peaks. When $K_1$ and $K_2$ are different by a order of magnitude, two peaks merge into one. Hence, the sensitivity is lowered.

\section{The Rotation Measurement for Dynamical Josephson Junctions\label{sec:tc}}
\begin{figure}[t]
\includegraphics[width=0.48\textwidth]{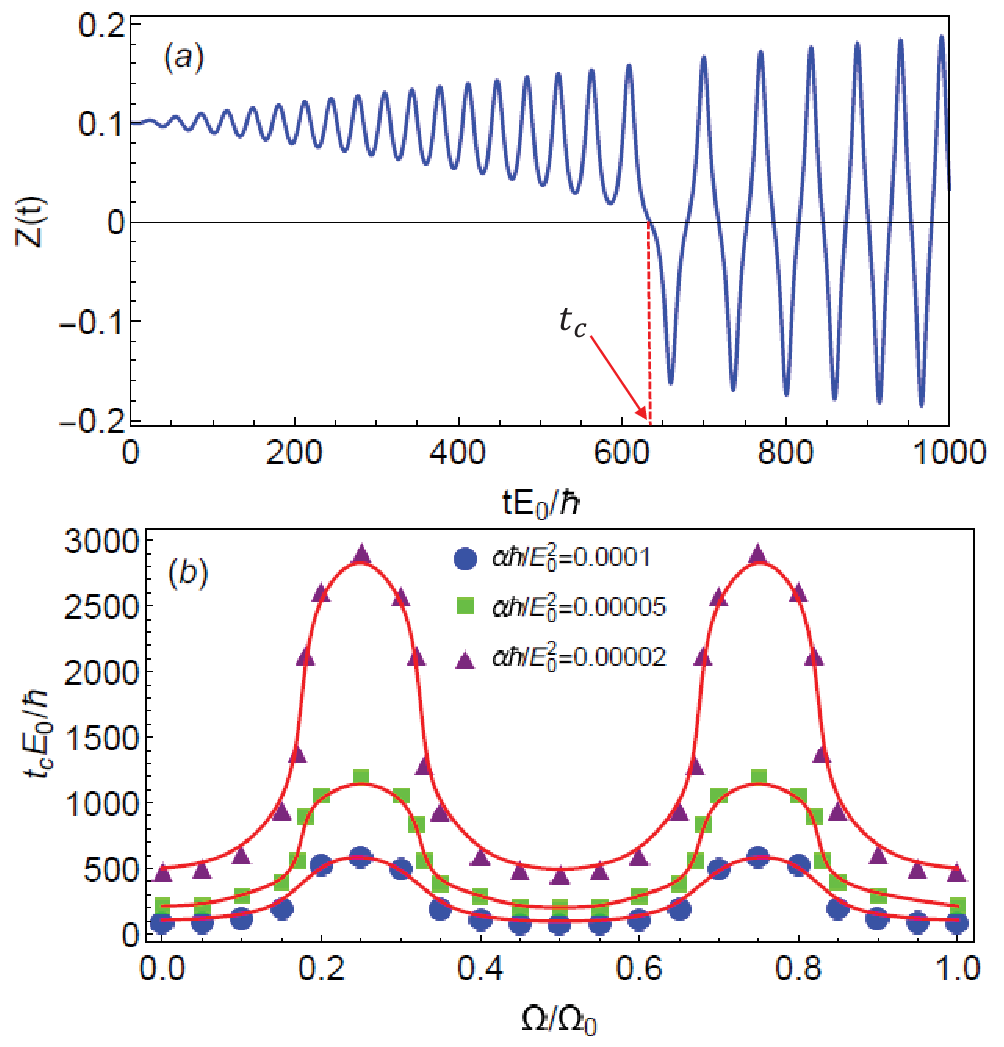}
\caption{(Color online) (a) The temporal variation of the population bias $Z(t)$ when the tunneling strength adiabatically changes as $K_1(t)=K_2(t)=\alpha t$, where $\alpha \hbar/E_0^2=10^{-5}$. (b) The critical time $t_c$ as a function of $\Omega/\Omega_0$ for different values of $\alpha$. Blue dots, green squares and brown triangles are for $\alpha \hbar/E_0^2=10^{-4}$, $\alpha \hbar/E_0^2=5\times10^{-5}$ and $\alpha \hbar/E_0^2=2\times10^{-5}$, respectively. The red solid lines are fitted by the method of cubic spline interpolation. The initial phase $\Phi(0)=\pi/2$ and  $\tilde U/E_0=1$ for all the graphs.}
\label{fig:VK}
\end{figure}

In this section we explore another physical quantity that responds to the rotation. As we have discussed in the previous section, the strength of the Josephson junctions can be easily tuned by changing the intensity of the laser beams. Here we consider dynamical Josephson junctions, that is, time dependent $K_1$ and $K_2$. For simplicity we take identical Josephson junctions and assume that they increase linearly with respect to time as $K_1(t)=K_2(t)=\alpha t$, where the parameter $\alpha$ is very small so that the changing of the system is an adiabatic process. The system is in the self-trapping regime initially since $K_1$ and $K_2$ are very small. As $K_1$ and $K_2$ increase to certain value the system changes from the self-trapping regime to the Josephson oscillation regime. Here we take the moment when $Z(t)$ reaches $0$ to characterize the transition as show in Fig. \ref{fig:VK} (a). We call it the critical time $t_c$. In \ref{fig:VK} (b) we plot the variation of the critical time $t_c$ with respect to the angular velocity $\Omega$ and find that $t_c$ also has a periodic modulation behavior. The period is $\Omega=0.5\Omega_0$, which is the same as the one of the critical population bias $Z_c$. Compared with the critical population bias the measurement of the critical time is an easier way to detect the rotation since the critical time can be find in one procedure of the experiment, while one has to take the experiment many times with different initial population bias $Z(0)$ to find the critical population bias $Z_c$.

\section{Conclusions\label{sec:con}}
In summary, we establish an analytic model for the ASQUID with two tunable junctions using tunneling Hamiltonian method. Based on this model we found two physical quantities that can be used for rotation sensing, the critical population bias $Z_c$ and the critical time $t_c$. We discussed the variation of $Z_c$ for two cases. First, when the junctions are symmetric, that is, $K_1=K_2$,  $Z_c$ demonstrates behavior of period modulation. The distance between two adjacent peaks is $\Omega_0/2$. Hence, the sensitivity of the ASQUID as a rotation sensor is determined by $\Omega_0$. In our model $\Omega_0=h/m\pi r_0^2$, hence the sensitivity is eventually determined by the mass of the atoms and the radius $r_0$ of the ring trap. Second, when the junctions are asymmetric, that is, $K_1\neq K_2$, we find that the asymmetry can lead to blurriness of the two adjacent peaks and thus lower the sensitivity. Finally we discussed the case of dynamic junctions, namely, the parameters $K_1$ and $K_2$ are time dependent. When $K_1$ and $K_2$ are adiabatically increasing the system can change from the self-trapping regime to the Josephson oscillation regime. The time of the transition is called the critical time $t_c$, which also demonstrates a behavior of periodic modulation due to the rotation. These investigations enrich the toolbox of motion sensing in cold atom physics.

Furthermore, the model and method we established in this work can be easily extended to more complicated systems, for instance, the ASQUIDs with spinor BECs or spin-orbit coupled superfluids, where we can explore more ways to detect the rotation.

\section{Acknowledgements}
The work is supported by the National Science Foundation of China (Grant No. NSFC-11874002).

\end{document}